\documentclass[aps,twocolumn,showpacs,preprintnumbers,amsmath,amssymb,superscriptaddress,floatfix,nofootinbib]{revtex4}
\usepackage{lipsum}
\usepackage{graphicx}
\usepackage{epsfig}
\usepackage{epstopdf}
\usepackage{hyperref}
\usepackage{amsmath}
\usepackage{amsfonts}
\usepackage{amssymb}
\usepackage{setspace}
\usepackage{amsfonts,color}

\begin{document}

\title{Semileptonic $\Lambda_b \to \bar \nu_l l \Lambda_c(2595)$ and $\Lambda_b \to \bar \nu_l l \Lambda_c(2625)$ decays in the molecular picture of $\Lambda_c(2595)$ and $\Lambda_c(2625)$ }

\author{Wei-Hong Liang}
\affiliation{Department of Physics, Guangxi Normal University,
Guilin 541004, China}
\affiliation{Departamento de
F\'{\i}sica Te\'orica and IFIC, Centro Mixto Universidad de
Valencia-CSIC Institutos de Investigaci\'on de Paterna, Aptdo.
22085, 46071 Valencia, Spain}

\author{Eulogio Oset}
 \affiliation{Departamento de
F\'{\i}sica Te\'orica and IFIC, Centro Mixto Universidad de
Valencia-CSIC Institutos de Investigaci\'on de Paterna, Aptdo.
22085, 46071 Valencia, Spain}

\author{Zhu-Sheng Xie}
\affiliation{Department of Physics, Guangxi Normal University,
Guilin 541004, China}

\date{\today}

\begin{abstract}
We evaluate the partial decay widths for the semileptonic $\Lambda_b \to \bar \nu_l l \Lambda_c(2595)$ and $\Lambda_b \to \bar \nu_l l \Lambda_c(2625)$ decays from the perspective that these two $\Lambda^*_c$ resonances are dynamically generated from the $DN$ and $D^*N$ interaction with coupled channels. We find that the ratio of the rates obtained for these two reactions is compatible with present experimental data and is very sensitive to the $D^* N$ coupling, which becomes essential to obtain agreement with experiment. Together with the results obtained for the $\Lambda_b \to \pi^- \Lambda^*_c$ reactions, it gives strong support to the molecular picture of the two $\Lambda^*_c$ resonances and the important role of the $D^*N$ component neglected in prior studies of the $\Lambda_c(2595)$ from the molecular perspective.
\end{abstract}

\maketitle

\section{Introduction}

The interaction of mesons with baryons using chiral dynamics and unitary in coupled channels, the chiral unitary approach \cite{weise,angels,ollerulf,Garcia,Hyodo} has brought light into the nature of some baryonic resonances. The prediction of two states for the $\Lambda (1405)$ \cite{ollerulf,cola} has been one example of it, and is now supported by experiments as shown in Refs.\,\cite{magas,seki} (see also note in the PDG concerning this issue \cite{notepdg}). In the charm sector the interaction of $DN$ and coupled channels has also been considered \cite{hofmann,mizutani} and, as a consequence, the $\Lambda_c(2595)$ resonance is generated dynamically, bearing many analogies to the $\Lambda(1405)$, one of which states couples strongly to $\bar KN$. While for some time only pseudoscalar-baryon channels were used, at some point it became clear that the mixture of pseudoscalar-baryon and vector-baryon should be relevant in some cases. A first step in this direction was given in Ref.\,\cite{garzon}, followed by Refs.\,\cite{kanchan,kanchan2} in the light sector and by Refs.\,\cite{uchino,uchino2} in the charm sector. Concerning the  $\Lambda_c(2595)$, the explicit consideration of the $DN$ and $D^*N$ channels, using pion exchange to connect them, is done in Ref.\,\cite{uchino}. In Refs.\,\cite{carmen,romanets}, SU(8) symmetry was used, with a symmetry breaking mechanism that gives rise to the Weinberg-Tomozawa interaction in the SU(3) sector. Both in Refs.\,\cite{uchino} and \cite{romanets}, it was found that the $\Lambda_c(2595)$ ($J^P=1/2^-$) couples strongly to $DN$ and $D^*N$ in $s$-wave. The $\Lambda_c(2625)$ ($J^P=3/2^-$) was also found dynamically generated, coupling strongly to $D^*N$ in  $s$-wave.

In a recent paper \cite{liangbayar}, the $\Lambda_b \to \pi^- \Lambda_c(2595)$ and  $\Lambda_b \to \pi^- \Lambda_c(2625)$ decays were studied, and it was found that they were very sensitive to the $DN$ and $D^*N$ couplings and to their relative sign. The experimental ratio of the branching fractions for the two decays was well reproduced with the results obtained in Ref.\,\cite{uchino}. It was found that the coupling of  $\Lambda_c(2595)$ to $D^*N$ was essential to obtain agreement with experiment, and if the relative sign of the couplings was reversed there was a cancellation of the $DN$ and $D^*N$ components that makes
the  $\Lambda_b \to \pi^- \Lambda_c(2595)$  partial decay width extremely small, in shear disagreement with experiment.

Support for the picture of Refs.\,\cite{uchino,romanets} should come from accumulation of experimental data which can be reproduced by the models. In this respect, in the present work we want to show one such reaction, the $\Lambda_b \to (\bar \nu_l l) \Lambda^*_c$, with $\Lambda^*_c \equiv \Lambda_c(2595), \Lambda_c(2625)$. We develop here the formalism that provides the width for these decays within the molecular picture of Ref.\,\cite{uchino} and show that the ratio of the branching fractions for these two reactions is compatible with present experimental data. Theoretical calculations for $\Lambda_b \to \bar \nu_l l \Lambda_c$, with $\Lambda_c$ the ground state, have been done before using constituent quark models \cite{albertus} and QCD lattice simulation \cite{juanito}. For $\Lambda^*_c$, this is the first calculation.

\section{Formalism}
\begin{figure}[htbp]\centering
\includegraphics[scale=0.65]{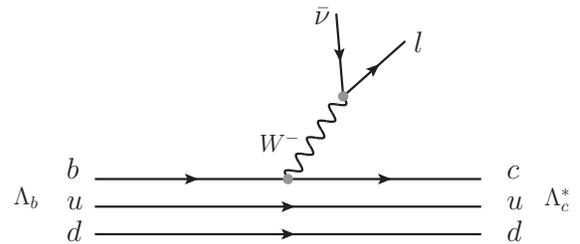}
\caption{Diagrammatic representation of the $\Lambda_b \to (l \bar \nu_l ) \Lambda^*_c$ decay.
\label{fig:FeynmanDiag1}}
\end{figure}

The picture for the $\Lambda_b \to (l \bar \nu_l ) \Lambda_c(2595)$ or $\Lambda_b \to (l \bar \nu_l ) \Lambda_c(2625)$ reactions
is given in Fig.\,\ref{fig:FeynmanDiag1} at the quark level.

We must bear in mind three important points:
1) The $ud$ quarks of the $\Lambda_b$ are in $I=0,\, S=0$ and they are spectators in the reaction. 2) Since in the final state, the $ud$ quarks still have $I=0,\, S=0$ and positive parity, the $c$ quark must carry negative parity to be able to produce the $1/2^-,\, 3/2^-$ $\Lambda^*_c$ states at the end. This means it will have $L=1$ in the quark picture. 3) Since the $\Lambda^*_c$ is generated from the $DN,\, D^* N$ interaction and other coupled channels, in the picture of  Fig.\,\ref{fig:FeynmanDiag1} one must include hadronization, creating a $\bar q q$ pair with the quantum numbers of the vacuum. The coupling with $\bar q q$ to give a meson-baryon system must include the $c$ quark to allow it to go back to the ground level, where it will be in the meson-baryon configuration.

The former considerations are similar to those done in the study of the $\Lambda_b \to J/\psi K^- p$ reaction studied in Ref.\,\cite{rocamai} and measured later in Ref.\,\cite{lhcbexp}. They were taken into account in the study of the $\Lambda_b \to \pi^- \Lambda^*_c$ decays in Ref.\,\cite{liangbayar} and we make use of the results here. It was found there that after taking into account the hadronization, including a singlet of SU(3) $\bar q q$ states $(\bar uu+\bar dd+\bar ss)$, the following hadronic configuration appeared at the end, ignoring the larger mass $D^+_s \Lambda$ component,
\begin{equation}\label{eq:Hprime}
   |H' \rangle = |D^0 p \rangle +|D^+ n \rangle \equiv \sqrt{2} \,|DN, I=0 \rangle.
 \end{equation}
Similarly, the same combination of $D^*N$ would appear, and the dynamics of the production of these two cases was explicitly studied in Ref.\,\cite{liangbayar}. We shall use some of the findings of that work here.

The dynamics in the present case is different than the one found in the $\pi^- \Lambda^*$ decay. As shown in Refs.\,\cite{sekihara,sekihara2}, the transition matrix is given by
\begin{equation}\label{eq:Tmatrix}
  T= -i G_F \frac{V_{bc}}{\sqrt{2}} L^{\alpha} \,Q_{\alpha} \,V_{\rm had},
\end{equation}
with $G_F$ the Fermi coupling constant, $V_{bc}$ the Cabibbo-Kobayashi-Maskawa matrix element for the $b \to c$ transition, $V_{\rm had}$ a factor accounting for the hadronic interaction, and $L^{\alpha}$, $Q_{\alpha}$ the leptonic and quark operators,
\begin{equation}\label{eq:LQoperators}
  L^\alpha \equiv \bar u_l \gamma^\alpha (1-\gamma_5) v_\nu, ~~~~Q_\alpha \equiv \bar u_c \gamma_\alpha (1-\gamma_5) u_b.
\end{equation}
When evaluating the sum and average over polarizations of $|T|^2$, we will have
\begin{equation}\label{eq:TSum}
  \frac{1}{2} \sum \sum |T|^2 \propto \frac{1}{2} \sum \sum \left| L^{\alpha} Q_{\alpha} \right|^2.
\end{equation}
As shown in Ref.\,\cite{sekihara}, we have
\begin{align}\label{eq:SumLL}
 \sum_{\rm pol} \! L^\alpha L^{\dag \beta} \!&=\! {\rm tr} \!\!\left[ \!\gamma^\alpha (1 \!-\!\gamma_5) \frac{p \!\!\!/ _\nu \!-\! m_\nu}{2m_\nu}  (1\!+\!\gamma_5) \gamma^\beta
  \frac{p \!\!\!/ _l \!+\!m_l}{2m_l}\right] \nonumber \\
 &= \!\!2\frac{p^\alpha _\nu p^\beta_l \!+\!p^\alpha _l p^\beta_\nu \!-\! p_\nu \! \cdot \! p_l g^{\alpha \beta}\!-\!i\epsilon^{\rho \alpha \sigma \beta} p_{\nu \rho} p_{l \sigma}}{m_\nu m_l}.
\end{align}

In Ref.\,\cite{sekihara}, a sum and average over polarization of the quarks was also done for $Q_{\alpha} Q^{\dag}_{\beta}$, but here we cannot do that if we want to differentiate between the production of $DN$ or $D^*N$. We divert from the formalism of Ref.\,\cite{sekihara} at this point, but recall from there that in the semileptonic processes the $\bar \nu_l l$ invariant mass is quite large, peaking around the end of the spectrum, which makes the $\Lambda^*$ come out with relatively small momentum, and, sharing this momentum with the $c,u,d$ quarks, the $c$ quark carries a small momentum at the end compared to its mass. Using the nonrelativistic expressions for $\gamma^{\mu}$ and $\gamma^{\mu} \gamma_5$ and neglecting terms that go like $p/m_c$, we find that only the $\gamma^0 \simeq 1$, and the $\gamma^i \gamma_5 \sim \sigma^i (i=1,2,3)$ components survive in this case. Then, after a bit of algebra, one easily finds
\begin{align}\label{eq:SumLLQQ}
 &\sum_{\rm lepton \, pol.} \!\!\!\!\!\!\!L^{\alpha} L^{\dag \beta}  Q_{\alpha} Q^{\dag}_{\beta}
   = \!\frac{2}{m_\nu m_l}
   \!\!\left[ 2p^0_\nu p^0_l -p_\nu \cdot p_l + 4p^0_\nu \vec p_l \cdot \vec \sigma \right. \nonumber \\
&+\!\! \left. (\vec p_\nu \cdot \vec \sigma)(\vec p_l \cdot \vec \sigma) +(\vec p_l \cdot \vec \sigma)(\vec p_\nu \cdot \vec \sigma) + (p_\nu \cdot p_l)(\vec \sigma \cdot \vec \sigma)
  \right],
\end{align}
where $\vec \sigma$ is acting at the level of quarks and the proper matrix elements with the quark polarizations, and sum and average over them, are still to be done.

\section{Quark matrix elements}
The $\vec \sigma$ operators in Eq.\,(\ref{eq:SumLLQQ}) act on the spins of the $b,c$ quarks, but, as mentioned above, the $c$ quark is in $L=1$. Then the quark matrix element that appears is
\begin{eqnarray}\label{eq:QME}
  \mathcal{M} &\equiv & \!\!\int d^3 r \varphi_{\rm in} (r) \varphi_{\rm out} (r) e^{-i\vec q \cdot \vec r} \,\sum_m \mathcal{C}(1\frac{1}{2}J;\, m, M'-m)\nonumber \\
   & \times &  \!Y^*_{1m}(\hat r)\, \langle \frac{1}{2}, M'-m \left| O_P \right| \frac{1}{2} M \rangle \, Y_{00} (\hat r),
\end{eqnarray}
where $Y^*_{1m}$ comes for the $c$ quark and $Y_{00} (\hat r)$ from the $b$ quark, and $e^{-i\vec q \cdot \vec r}$ stands for the plane wave of the $\bar \nu_l l$ emitted pair with momentum $\vec q$. In Eq.\,(\ref{eq:QME}), $J$ is the total angular momentum of the $c$ quark, which coincides with the spin of the $\Lambda_c^*$, $1/2$ for $\Lambda_c(2595)$ and $3/2$ for $\Lambda_c(2625)$, since the $ud$ pair and the $\bar q q$ carry both $J'=0$. The operator $O_P$ will be $1$ or $\vec \sigma$ depending on the terms in Eq.\,(\ref{eq:SumLLQQ}). Expanding $e^{-i\vec q \cdot \vec r}$ in partial waves, we have
\begin{equation}\label{eq:exp}
 e^{-i\vec q \cdot \vec r}=4\pi \sum_l (-i)^l j_l(qr) \sum_\mu (-1)^\mu \,Y^*_{l\mu}(\hat r)\, Y^*_{l,-\mu}(\hat q).
\end{equation}
After performing the $d\Omega (\hat r)$ integration we get
\begin{eqnarray}\label{eq:QME2}
  \mathcal{M} &= & -4\pi i \sum_m \mathcal{C}(1\frac{1}{2}J;\, m, M'-m)\, Y^*_{1m}(\hat q) \nonumber \\
  &&\times \langle \frac{1}{2}, M'-m \left| O_P \right| \frac{1}{2}, M \rangle \, ME(q) \,Y_{00},
\end{eqnarray}
with
\begin{equation}\label{eq:MEq}
  ME(q)\equiv \int r^2 dr  \varphi_{\rm in} (r)\, \varphi_{\rm out} (r) \, j_1 (qr).
\end{equation}

In Ref.\,\cite{liangbayar}, the matrix elements $\mathcal{M}$ were written in terms of the macroscopic $\vec \sigma$ and $\vec S^+$ operators acting on the $\Lambda_b$ and $\Lambda^*_c$, where $\vec S^+$ is the transition spin operator from spin $1/2$ to $3/2$, normalized such that
\begin{equation}\label{eq:SOperator}
  \sum_M S_i | M \rangle \langle M | S^+_j =\frac{2}{3} \delta_{ij} -\frac{i}{3} \epsilon_{ijk} \sigma_k.
\end{equation}
The results obtained are summarized in Table \ref{tab:O_P}, omitting the $ME(q)$ factor.
\begin{table}[htbp]
     \renewcommand{\arraystretch}{1.5}
     \setlength{\tabcolsep}{0.4cm}
\caption{\label{tab:O_P} Macroscopic operators in the $\Lambda_b \to \Lambda^*_c$ transitions associated to the microscopic operators at quark level.}
\centering
\begin{tabular}{r|cc}
\hline \hline
 & $O_P=1$  &  $O_P= \vec \sigma \cdot \vec q$  (quark level)  \\
\hline
 $J=1/2$ & $ i \frac{\vec \sigma \cdot \vec q}{q}$   & $iq$   \\
$J=3/2$  &$ -i \sqrt{3}\frac{\vec S^+ \cdot \vec q}{q}$  & 0 \\
\hline \hline
\end{tabular}
\end{table}

Coming back to Eq.\,(\ref{eq:SumLLQQ}), let us evaluate these terms now. We can take advantage that for the $\Lambda_b$ the spin of the $b$ quark is the same as the one of the $\Lambda_b$, since the $ud$ pair comes $S=0$. Then, in the sum over the third component of the $\Lambda_b$ spin, $M$, we would have
\begin{equation}\label{eq:sigmaOperator}
  \sum_M \sigma^i \, |\frac{1}{2} M \rangle \langle \frac{1}{2} M | \, \sigma^i \equiv \delta_{ii} +i \epsilon_{ijk} \sigma_k =3,
\end{equation}
and
\begin{eqnarray}
  &&\sum_M \left[ p_{\nu_i} \sigma_i \, |\frac{1}{2} M \rangle \langle \frac{1}{2} M | \, p_{l_j} \sigma_j
  \right. \nonumber \\
  &&~~~~\left.+p_{l_i} \sigma_i \, |\frac{1}{2} M \rangle \langle \frac{1}{2} M | \, p_{\nu_j} \sigma_j \right] \nonumber \\
  &=& \left(  p_{\nu_i} p_{l_j} +p_{\nu_j} p_{l_i} \right) \left( \delta_{ij} + i \epsilon_{ijk} \sigma_k \right) \nonumber \\
  &=& 2 \vec p_{\nu} \cdot \vec p_{l}.
\end{eqnarray}
On the other hand, we can write the term $\vec p_{l} \cdot \vec \sigma$ as
\begin{equation*}
  \vec p_{l} \cdot \vec \sigma = \frac{1}{2} \left(  \vec q + \vec p_{r} \right)\cdot \vec \sigma
\end{equation*}
where
\begin{equation*}
  \vec q = \vec p_{l} + \vec p_{r}, ~~~~ \vec p_{r} = \vec p_{l} - \vec p_{\nu}.
\end{equation*}
Now, according to Table \ref{tab:O_P},
the term $p^0_\nu \,\vec \sigma \cdot \vec q$ at the quark level, containing the operators $1$ and $\vec \sigma \cdot \vec q$,  will give rise to $p^0_\nu \,i \frac{\vec \sigma \cdot \vec q}{q} (-)iq$ at the macroscopic level for $J=1/2$ and zero for $J=3/2$.
But the trace of $\vec \sigma \cdot \vec q$, when summing over polarizations will be zero. The term $\vec p_r \cdot \vec \sigma$ will also vanish when one integrates over the angles, as we shall see in the next section. Hence, this term vanishes in the sum over polarizations and integration over the phase space. We are thus left with
\begin{eqnarray}\label{eq:SumLLQQ2}
   && \sum_{\rm lepton \, pol.} \!\!\!L^{\alpha} L^{\dag \beta}  Q_{\alpha} Q^{\dag}_{\beta}\nonumber \\
    &= &
    \frac{2}{m_\nu m_l} \left[ 2 p^0_\nu p^0_l - p_\nu \cdot p_l + 2 \vec p_\nu \cdot \vec p_l + 3 p_\nu \cdot p_l \right] \nonumber \\
 &=& \frac{8}{m_\nu m_l} p^0_\nu p^0_l.
 \end{eqnarray}
Now if we want to evaluate the extra sum over polarizations of $\Lambda_b$ and $\Lambda^*_c$, we go to the macroscopic representation of Table \ref{tab:O_P} and have
\begin{widetext}
 \begin{spacing}{2.0}
\begin{equation}\label{eq:SumLLQQ3}
  \overline{\sum}\sum L^{\alpha} L^{\dag \beta}  Q_{\alpha} Q^{\dag}_{\beta}
    \equiv \frac{8}{m_\nu m_l}\, p^0_\nu \, p^0_l \,\frac{1}{2\vec q^2}\sum_{M,M'} \left\{
    \begin{array}{ll}
     \langle \frac{1}{2} M \left| \right. \vec \sigma \cdot \vec q \left. \right| \frac{1}{2} M' \rangle
    \langle \frac{1}{2} M' \left|  \right. \vec \sigma \cdot \vec q \left. \right| \frac{1}{2} M \rangle,& {\rm for~} J=1/2; \\
    3 \langle \frac{1}{2} M \left| \right. \vec S \cdot \vec q \left. \right| \frac{3}{2} M' \rangle
    \langle \frac{3}{2} M' \left| \right. \vec S^+ \cdot \vec q \left. \right| \frac{1}{2} M \rangle,& {\rm for~} J=3/2. \\
    \end{array}
   \right.
\end{equation}
 \end{spacing}
\end{widetext}
Hence,
\begin{equation}\label{eq:SumLLQQ4}
  \overline{\sum}\sum L^{\alpha} L^{\dag \beta}  Q_{\alpha} Q^{\dag}_{\beta}
    \equiv A_J \, \frac{8}{m_\nu m_l} p^0_\nu p^0_l,
\end{equation}
with $A_{1/2}=1$ and $A_{3/2}=2$.

\begin{figure}[tb]\centering
\includegraphics[scale=0.55]{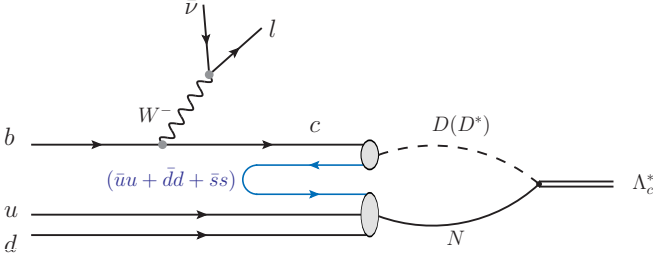}
\caption{Mechanism to produce a $\Lambda^*_c$ resonance that is mostly made from $DN, D^*N$.
\label{fig:FeynmanDiag2}}
\end{figure}

There is still one more element to consider, which is to include the molecular dynamics of the $\Lambda^*_c$ states. To connect to the $DN$ and $D^*N$ components one must take into account the hadronization of the $\bar qq$ pair. This was done in Ref.\,\cite{liangbayar}. The mechanism is depicted
in Fig.\,\ref{fig:FeynmanDiag2} and introduces a hadronic factor different for $DN$ and $D^*N$ coupling to $J=1/2$, and for $D^*N$ coupling to $J=3/2$.
The result of the hadronic factor $V_{\rm had}$ for these cases is given in Table \ref{tab:Vhad}, up to a common global factor, where $G_{D N}$ and $G_{D^* N}$ are the $DN (D^*N)$ loop functions for the propagation of these states in Fig.\,\ref{fig:FeynmanDiag2}, and $g_{R,DN}$ and $g_{R,D^*N}$ the couplings of the $\Lambda_c(2595)$,  $\Lambda_c(2625)$ to the $DN$ and $D^*N$ states.
All this information is given in Ref.\,\cite{uchino} and we summarize it in Table \ref{tab:Ggcoupling}.
\begin{table}[htbp]
     \renewcommand{\arraystretch}{1.5}
     \setlength{\tabcolsep}{0.4cm}
\caption{\label{tab:Vhad} Contributions to $V_{\rm had}$ from $DN$ and $D^*N$ in the coupling to $J=1/2$ and $J=3/2$ from Ref.\,\cite{liangbayar}.
The couplings $g_i$ and the loop functions $G_i$ are obtained in
Ref.\,\cite{uchino}.\protect\footnote{Note change of sign in the $D^*N$ case, as discussed in Ref.\,\cite{liangbayar}, because in Ref.\,\cite{uchino} $V^2_{\rm eff}$ for the $DN \to D^*N$ transition was calculated and the positive sign for $V_{\rm eff}$ was taken by default. The right sign, corresponding to $\pi$ exchange, is negative. This sign is not relevant for the spectrum discussed in Ref.\,\cite{uchino}, but it matters here.}}
\centering
\begin{tabular}{r|cc}
\hline \hline
$V_{\rm had}$ & $DN$  &  $D^*N$    \\
\hline
 $J=1/2$ & $ \frac{1}{2} G_{DN} \cdot g_{R,DN}$   & $\frac{1}{2\sqrt{3}} G_{D^*N}\cdot g_{R,D^*N}$   \\
$J=3/2$  &$ 0$  & $\frac{1}{\sqrt{3}} G_{D^*N}\cdot g_{R,D^*N}$ \\
\hline \hline
\end{tabular}
\end{table}

\begin{table}[htbp]
     \renewcommand{\arraystretch}{1.5}
     \setlength{\tabcolsep}{0.4cm}
\caption{\label{tab:Ggcoupling} The values of $G_{DN}\cdot g_{R,DN}$ and $G_{D^*N} \cdot g_{R,D^*N}$ for the two $\Lambda^*_c$ resonances.}
\centering
\begin{tabular}{r|cc}
\hline \hline
& $G_{DN}\cdot g_{R,DN}$  &  $G_{D^*N} \cdot g_{R,D^*N}$   \\
\hline
$\Lambda_c(2595)$  & $13.88 - 1.06i$ & $26.51 + 2.1i$   \\
$\Lambda_c(2625)$ & $0$  & $29.10$ \\
\hline \hline
\end{tabular}
\end{table}

With all these ingredients, we can write
\begin{equation}\label{eq:Tsqure}
  \overline{\sum} \sum |T|^2 =C \frac{8}{m_\nu m_l} p^0_\nu p^0_l \, A_J V_{\rm had} (J),
\end{equation}
\begin{align}\label{eq:AJVhad}
   & A_J V_{\rm had} (J) \nonumber\\
  \equiv & \left\{
  \begin{array}{ll}
    \left|  \frac{1}{2} G_{DN} \cdot g_{R,DN} + \frac{1}{2\sqrt{3}} G_{D^*N} \cdot g_{R,D^*N}  \right|^2, & {\rm for~} J=1/2 \\[3mm]
    2\left| \frac{1}{\sqrt{3}}  G_{D^*N} \cdot g_{R,D^*N}  \right|^2, & {\rm for~} J=3/2\\
    \end{array}
   \right.
\end{align}
where $C$ is a global common factor that contains $ME(q)^2$. With values of $M_{\rm inv} (\bar \nu_l l)$ large, the values of $q$ are not large. We can consider it a smooth function over the phase space. However, since the only observable that we want to evaluate is the ratio of branching fractions, this ratio is essentially given by the ratio of $A_J V_{\rm had} (J)$ for the two resonances since the integrals of phase space are practically identical for the two resonances. Before we perform the numerical calculations of the phase space, we can already quote here that
\begin{equation}\label{eq:AVRatio}
  \frac{A_{1/2} \, V_{\rm had} (1/2)}{A_{3/2} \, V_{\rm had} (3/2)} = 0.38,
\end{equation}
and this should be very similar to the final result considering the slight differences in the phase space.

\section{Evaluation of the width}
The width for the decay into three particles is given by
\begin{align}\label{eq:Width}
  \Gamma \! &=\! 2M_{\Lambda^*_c} 2m_{\nu} 2m_l \frac{1}{2} \frac{1}{(2\pi)^3} \frac{1}{M_{\Lambda_b}} \nonumber\\
   &\! \times \!\!\int {\rm d} M_{\rm inv}(\bar \nu l) p_{\Lambda^*_c} {\rm d} \Omega_{\Lambda^*_c} \int d \tilde{\Omega}_l \frac{1}{16 \pi^2} \tilde{p}_l \overline{\sum} \sum |T|^2,
\end{align}
where $p_{\Lambda^*_c}=-\vec q$ is the momentum of the $\Lambda^*_c$ in the rest frame of $\Lambda_b$, and $\tilde{p}_l$ the lepton momentum in the $\bar \nu l$ rest frame, respectively. One can see that the masses of the lepton and neutrino, which appear because of our choice of normalization of the fermion field, cancel in the final expression of Eq.\,(\ref{eq:Width}).
Once we arrive to this point we can come back to see why the $\vec \sigma \cdot \vec p_r$ term vanishes in the phase space integration. To see that, it is interesting to make a boost from the $\bar \nu l$ rest frame to the $\Lambda_b$ rest frame where the $\bar \nu l$ pair has an energy $E_{\nu l}$ and a momentum $\vec q$. We obtain
\begin{equation}\label{eq:pl}
  \vec p_l = \left[  \left( \frac{E_{\nu l}}{M_{\rm inv}}-1 \right) \frac{\vec{\tilde{p}}_l \cdot \vec q}{\vec q^2} + \frac{\tilde{p}^0_l}{M_{\rm inv}}\right] \vec q +\vec{\tilde{p}}_l,
\end{equation}
where $\vec p_l$ and $\vec{\tilde{p}}_l$ are the lepton momenta in the $\Lambda_b$ rest frame and in the $\bar \nu l$ rest frame, respectively, and $\tilde{p}^0_l$ the lepton energy in the $\bar \nu l$ rest frame. Since $\vec{\tilde{p}}_\nu=-\vec{\tilde{p}}_l$, then
\begin{equation}\label{eq:pr}
  \vec p_r = \vec p_l -\vec p_\nu =2 \vec{\tilde{p}}_l + 2 \left( \frac{E_{\nu l}}{M_{\rm inv}}-1 \right) \frac{\vec{\tilde{p}}_l \cdot \vec q}{\vec q^2} \vec q,
\end{equation}
where we have considered that $\tilde{E}_\nu = \tilde{E}_l$, assuming zero mass for both of them.

We can see in Eq.\,(\ref{eq:pr}) that when integrating over ${\rm d} \tilde{\Omega}_l$ the vector $\vec p_r$, proportional to $\vec{\tilde{p}}_l$, will vanish in the integration.

One last point is a practical one to reduce the integral of $\Gamma$ in Eq.\,(\ref{eq:Width}) to just one numerical integration. For this we follow the steps of Ref.\,\cite{sekihara}.

The factor $p^0_\nu p^0_l$ in Eq.\,(\ref{eq:SumLLQQ4}) evaluated in the $\Lambda_b$ rest frame, where we could reduce the $\gamma^\mu, \gamma^\mu \gamma_5$ matrices to easy expressions, will depend on angles and we should in principle perform all the integrals in Eq.\,(\ref{eq:Width}). We can write in an invariant way
\begin{equation}\label{eq:p0}
  p^0_\nu p^0_l \, (\Lambda_b {\rm ~rest ~ frame})= \frac{1}{M^2_{\Lambda_b}} (p_\nu \cdot p_{\Lambda_b}) (p_l \cdot p_{\Lambda_b} ).
\end{equation}

Next we evaluate these invariant products in the frame where $\bar \nu l$ is at rest. In this frame, $\vec{\tilde{p}}_l =- \vec{\tilde{p}}_\nu$, $\vec{\tilde{p}}_{\Lambda_b} = \vec{\tilde{p}}_{\Lambda^*_c}$ and using $\tilde{E}_{\Lambda_b} =M_{\rm inv} + \tilde{E}_{\Lambda^*_c}$ we obtain
\begin{equation*}
  \tilde{E}_{\Lambda_b} \!=\!\frac{M^2_{\Lambda_b} \!+\! M^2_{\rm inv} \!-\!M^2_{\Lambda^*_c}}{2 M_{\rm inv}},~~~
   |\vec{\tilde{p}}_{\Lambda_b}| \!=\! \frac{\lambda^{1/2} (M^2_{\Lambda_b}, M^2_{\rm inv}, M^2_{\Lambda^*_c})}{2 M_{\rm inv}}.
\end{equation*}
Then,
\begin{align}\label{eq:pppp}
   & \frac{1}{M^2_{\Lambda_b}}  (p_\nu \cdot p_{\Lambda_b}) (p_l \cdot p_{\Lambda_b} ) \nonumber \\
  = & \frac{1}{M^2_{\Lambda_b}} \left[ \left(  \tilde{p}^0_{\nu} \tilde{E}_{\Lambda_b} - \vec{\tilde{p}}_\nu \cdot \vec{\tilde{p}}_{\Lambda_b}  \right)
  \left( \tilde{p}^0_l \tilde{E}_{\Lambda_b} - \vec{\tilde{p}}_l \cdot \vec{\tilde{p}}_{\Lambda_b}  \right) \right] \nonumber \\
 = & \frac{1}{M^2_{\Lambda_b}} \left[ \left( \tilde{p}^0_{\nu} \tilde{E}_{\Lambda_b}  \right)^2  -\left( \vec{\tilde{p}}_\nu \cdot \vec{\tilde{p}}_{\Lambda_b}  \right)^2 \right].
\end{align}
Taking into account that
\begin{equation*}
  \int {\rm d} \tilde{\Omega}_l |\vec{\tilde{p}}_l| |\vec{\tilde{p}}_{\Lambda_b}|^2 \cos^2 \theta = \frac{1}{3} \int {\rm d} \tilde{\Omega}_l |\vec{\tilde{p}}_l| |\vec{\tilde{p}}_{\Lambda_b}|^2,
\end{equation*}
and that
\begin{equation*}
  \tilde{p}^0_\nu = \frac{M_{\rm inv}}{2} =|\vec{\tilde{p}}_l| ,
\end{equation*}
we obtain, that we can neglect the angle dependence of $p^0_\nu p^0_l$ and use over the whole phase space
\begin{equation}\label{eq:p0p0}
  p^0_\nu p^0_l \rightarrow \frac{1}{M^2_{\Lambda_b}} \left( \frac{M_{\rm inv}}{2} \right)^2 \left[ \tilde{E}^2_{\Lambda_b} - \frac{1}{3} \vec{\tilde{p}}^{\:2}_{\Lambda_b}\right],
\end{equation}
 and the width is now given by
 \begin{equation}\label{eq:width2}
   \Gamma = \int \frac{{\rm d} \Gamma}{{\rm d} M_{\rm inv}} {\rm d} M_{\rm inv}
 \end{equation}
with
\begin{equation}\label{eq:dGamma}
  \frac{{\rm d} \Gamma}{{\rm d} M_{\rm inv}} = 2M_{\Lambda_b} 2M_{\Lambda^*_c} 2m_\nu 2m_l \frac{1}{4M^2_{\Lambda_b}} \frac{1}{(2\pi)^3} p_{\Lambda^*_c} \tilde{p}_l \overline{\sum} \sum |T|^2,
\end{equation}
where in $|T|^2$ of Eq.\,(\ref{eq:Tsqure}) we substitute $ p^0_\nu p^0_l$ by the expression of Eq.\,(\ref{eq:p0p0}), and $p_{\Lambda^*_c}$, $\tilde{p}_l$ are given by
\begin{eqnarray}\label{eq:pLamc}
  p_{\Lambda^*_c} &=& \frac{\lambda^{1/2} (M^2_{\Lambda_b}, M^2_{\rm inv}, M^2_{\Lambda^*_c})}{2 M_{\Lambda_b}},  \nonumber \\
  |\tilde{p}_l| &=& \frac{\lambda^{1/2} ( M^2_{\rm inv}, m^2_l, m^2_\nu)}{2 M_{\rm inv}}\equiv\frac{M_{\rm inv}}{2}.\nonumber
\end{eqnarray}

\section{Results}
\begin{figure}[tb]\centering
\includegraphics[scale=0.25]{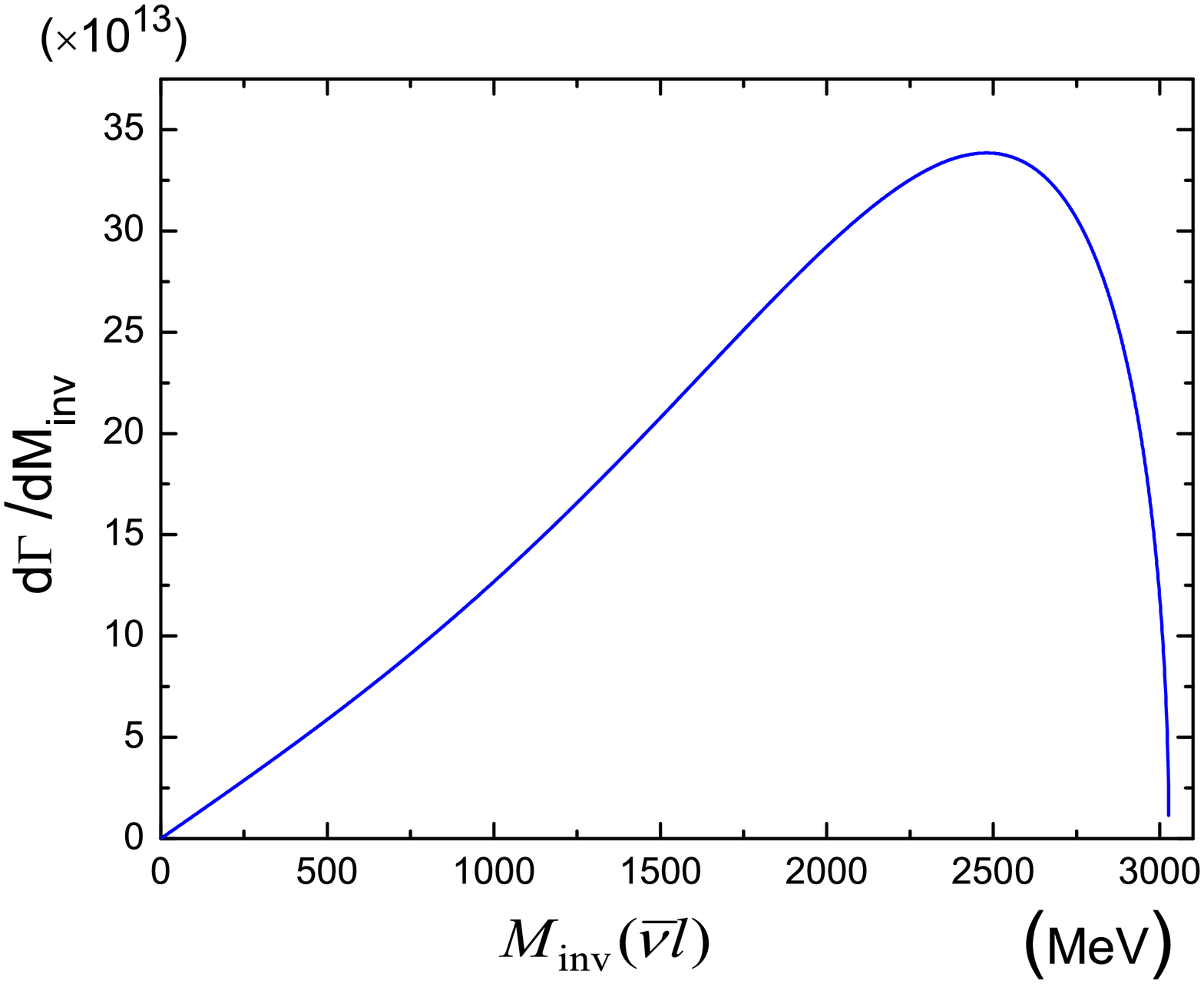}
\caption{$\frac{{\rm d}\Gamma}{{\rm d} M_{\rm inv}}$ for the $(\bar \nu l)$ pair as a function of $M_{\rm inv} (\bar \nu l)$ in the $\Lambda_b \to \bar \nu_l l \Lambda_c(2595)$ decay.
\label{fig:Minv}}
\end{figure}

We evaluate first $\frac{{\rm d}\Gamma}{{\rm d} M_{\rm inv}}$ for the $\Lambda_c(2595)$ by means of Eq.\,(\ref{eq:dGamma}), and the results are shown in Fig.\,\ref{fig:Minv}.
The result for $\Lambda_c(2625)$ has a nearly identical shape. As we can see from Fig.\,\ref{fig:Minv}, the $(\bar \nu l)$ invariant mass distribution peaks around the end of the phase space. Thanks to this, the momenta of $\Lambda^*_c$ are relatively small, justifying the nonrelativistic approximations made in $\gamma^\mu$ and $\gamma^\mu \gamma_5$ in the $Wbc$ vertex.

On the other hand, by taking $C$ constant in Eq.\,(\ref{eq:Tsqure}) and using Eqs.\,(\ref{eq:Tsqure}),(\ref{eq:p0p0}),(\ref{eq:dGamma}), we evaluate $\Gamma [\Lambda_b \to \bar \nu_l l \Lambda_c(2595)]$ and $\Gamma [\Lambda_b \to \bar \nu_l l \Lambda_c(2625)]$. We eliminate $C$ by taking the ratio of the two widths, and we find
\begin{equation}\label{eq:GammaVSGamma}
  \frac{\Gamma [\Lambda_b \to \bar \nu_l l \Lambda_c(2595)]}{\Gamma [\Lambda_b \to \bar \nu_l l \Lambda_c(2625)]} =0.39.
\end{equation}
As we can see, this result is practically identical to the one obtained in Eq.\,(\ref{eq:AVRatio}). The effect of considering the different phase space in the two reactions of Eq.\,(\ref{eq:GammaVSGamma}) is an increase of the ratio by $3\%$ with respect to the result obtained in Eq.\,(\ref{eq:AVRatio}).

The experimental data from the PDG are \cite{pdg}
\begin{eqnarray}\label{eq:BR_PDG}
  BR[\Lambda_b \to \bar \nu_l l \Lambda_c(2595)] &=& \left( 7.9 ^{+4.0}_{-3.5} \right) \times 10^{-3}, \nonumber \\
  BR[\Lambda_b \to \bar \nu_l l \Lambda_c(2625)] &=& \left( 13.0 ^{+6.0}_{-5.0} \right) \times 10^{-3}.
\end{eqnarray}
The ratio, summing in quadrature the experimental errors is
\begin{equation}\label{eq:GammaVSGamma_Exp}
  \left.\frac{\Gamma [\Lambda_b \to \bar \nu_l l \Lambda_c(2595)]}{\Gamma [\Lambda_b \to \bar \nu_l l \Lambda_c(2625)]} \right|_{\rm Exp.} =0.6^{+0.4}_{-0.3}.
\end{equation}
We can see that there is agreement between theory and experiment within errors.

The agreement obtained is not trivial and essentially tied to the $D^*N$ component of the $\Lambda^*_c(2595)$ resonance. Should there be no coupling to $D^*N$, we would have obtained a ratio for Eq.\,(\ref{eq:GammaVSGamma}) of the order of $0.1$, clearly in contradiction with experiment, even within the large errors. On the other hand, should the relative sign between $g_{R,D^*N}$ and $g_{R,DN}$ be the opposite, we would have obtained a ratio for Eq.\,(\ref{eq:GammaVSGamma}) of $0.02$ in shear contradiction with experiment.

The reactions studied and their ratio of widths give support to the molecular picture of the $\Lambda_c(2595)$ and $\Lambda_c(2625)$ as dynamically generated from $DN, D^*N$ and other coupled channels, described in Ref.\,\cite{uchino}, with $DN$ and $D^*N$ as the more important components. Together with the results obtained in Ref.\,\cite{liangbayar} for the $\Lambda_b \to \pi^- \Lambda_c(2595)$ and  $\Lambda_b \to \pi^- \Lambda_c(2625)$, they provide a boost to this molecular picture. It would be good to have evaluations of these ratios with different pictures, as well as have more experiments with the production of these resonances which can be contrasted with the different pictures.

\section{Conclusions}
We have studied the $\Lambda_b \to \bar \nu_l l \Lambda_c(2595)$ and $\Lambda_b \to \bar \nu_l l \Lambda_c(2625)$ reactions from the perspective that the two $\Lambda^*_c$ resonances are dynamically generated from the  $DN, D^*N$ interaction with coupled channels. We work out microscopically the weak vertices, involving $Wbc$ and $W\nu l$, to have a $DN, D^*N$ baryonic final state, which couples to $\Lambda^*_c$. For this, a $\bar q q$ pair with the quantum numbers of the vacuum  is created and the $c\bar q$ combine to give either the $D$ or $D^*$. With the help of Racah algebra, one can work out the weight for the formation of the $DN$ and $D^* N$ components. This, together with the coupling of $DN, D^*N$ to the two $\Lambda^*_c$ states, gives finally the amplitudes for the
$\Lambda_b \to \bar \nu_l l \Lambda^*_c$ transitions. With the input for the $DN, D^*N$ couplings to $\Lambda^*_c$ obtained in Ref.\,\cite{uchino} we can evaluate the rates for these transitions, up to a common factor involving radial matrix elements of the $b$ and $c$ wave functions. We do not evaluate this matrix element, which involves an excited $c$ quark state in $L=1$, but calculate the ratio of partial decay widths, where this factor cancels. We obtain results which are in agreement with experiment, within errors, and note that the agreement is obtained thanks to the coupling of the $\Lambda_c(2595)$ to the $D^*N$ component, which was neglected in early studies of this resonance. This agreement adds to the one found before for the  $\Lambda_b \to \pi^- \Lambda^*_c$ reactions. One should note that the ratio found for these latter branching fractions ($\Lambda_c(2595)$ versus $\Lambda_c(2625)$) was about $0.74$, while the one for the semileptonic reactions has been found of the order of $0.4$. The experimental data also follow this trend, the ratio for $\pi^- \Lambda^*_c$ secays is of the order of $1.0 \pm 0.6$, while the one of the semileptonic decays is about $0.6 \pm 0.4$. The relative weight of these ratios for the central values is very similar in the theory and the experiment.

The results obtained with these reactions give support to the molecular picture for these two $\Lambda^*_c$ resonances. Work with other models and checks for further experiments will help us gain further insight on the nature of these resonances, and new experiments producing these two resonances should be encouraged.

\section*{Acknowledgments}

This work is partly supported by the
National Natural Science Foundation of China under Grants
No.~11565007 and No.~11547307.
This work is also partly supported by the Spanish Ministerio
de Economia y Competitividad and European FEDER funds
under the contract number FIS2011-28853-C02-01, FIS2011-
28853-C02-02, FIS2014-57026-REDT, FIS2014-51948-C2-
1-P, and FIS2014-51948-C2-2-P, and the Generalitat Valenciana
in the program Prometeo II-2014/068.


\bibliographystyle{plain}

\begin{thebibliography}{999}


\bibitem{weise}
  N.~Kaiser, P.~B.~Siegel and W.~Weise,
  Nucl.\ Phys.\ A {\bf 594}, 325 (1995)
  [nucl-th/9505043].


\bibitem{angels}
 E.~Oset and A.~Ramos,
  Nucl.\ Phys.\ A {\bf 635}, 99 (1998)
  [nucl-th/9711022].


\bibitem{ollerulf}
 J.~A.~Oller and Ulf-G. Mei{\ss}ner,
  Phys.\ Lett.\ B {\bf 500}, 263 (2001)
  [hep-ph/0011146].


\bibitem{Garcia}
 C.~Garcia-Recio, J.~Nieves, E.~Ruiz Arriola and M.~J.~Vicente Vacas,
  Phys.\ Rev.\ D {\bf 67}, 076009 (2003)
  [hep-ph/0210311].


\bibitem{Hyodo}
 T.~Hyodo and D.~Jido,
  Prog.\ Part.\ Nucl.\ Phys.\  {\bf 67}, 55 (2012)
  [arXiv:1104.4474 [nucl-th]].

\bibitem{cola}
 D.~Jido, J.~A.~Oller, E.~Oset, A.~Ramos and Ulf-G. Mei{\ss}ner,
  Nucl.\ Phys.\ A {\bf 725}, 181 (2003)
  [nucl-th/0303062].


\bibitem{magas}
 V.~K.~Magas, E.~Oset and A.~Ramos,
  Phys.\ Rev.\ Lett.\  {\bf 95}, 052301 (2005)
  [hep-ph/0503043].

\bibitem{seki}
 D.~Jido, E.~Oset and T.~Sekihara,
  Eur.\ Phys.\ J.\ A {\bf 42}, 257 (2009)
  [arXiv:0904.3410 [nucl-th]].


\bibitem{notepdg}
Review on "Pole Structure of the $\Lambda(1405)$ Region", by  Ulf-G. Mei{\ss}ner, T. Hyodo;
Review of Particle Physics, C. Patrignani et al. (Particle Data Group), Chin. Phys. C, {\bf 40}, 100001 (2016).


\bibitem{hofmann}
J.~Hofmann and M.~F.~M.~Lutz,
  Nucl.\ Phys.\ A {\bf 763}, 90 (2005)
  [hep-ph/0507071].

\bibitem{mizutani}
T.~Mizutani and A.~Ramos,
  Phys.\ Rev.\ C {\bf 74}, 065201 (2006)
  [hep-ph/0607257].


\bibitem{garzon}
 E.~J.~Garzon and E.~Oset,
  Eur.\ Phys.\ J.\ A {\bf 48}, 5 (2012)
  [arXiv:1201.3756 [hep-ph]].


\bibitem{kanchan}
K.~P.~Khemchandani, A.~Martinez Torres, H.~Nagahiro and A.~Hosaka,
  Nucl.\ Phys.\ A {\bf 914}, 300 (2013).


\bibitem{kanchan2}
 K.~P.~Khemchandani, A.~Martinez Torres, F.~S.~Navarra, M.~Nielsen and L.~Tolos,
  Phys.\ Rev.\ D {\bf 91}, 094008 (2015)
  [arXiv:1406.7203 [nucl-th]].


\bibitem{uchino}
 W.~H.~Liang, T.~Uchino, C.~W.~Xiao and E.~Oset,
  Eur.\ Phys.\ J.\ A {\bf 51}, no. 2, 16 (2015)
  [arXiv:1402.5293 [hep-ph]].


\bibitem{uchino2}
 T.~Uchino, W.~H.~Liang and E.~Oset,
  Eur.\ Phys.\ J.\ A {\bf 52}, no. 3, 43 (2016)
  [arXiv:1504.05726 [hep-ph]].

\bibitem{carmen}
  C.~Albertus, E.~Hernandez and J.~Nieves,
  Phys.\ Rev.\ D {\bf 71}, 014012 (2005)
  [nucl-th/0412006].

\bibitem{romanets}
 O.~Romanets, L.~Tolos, C.~Garcia-Recio, J.~Nieves, L.~L.~Salcedo and R.~G.~E.~Timmermans,
  Phys.\ Rev.\ D {\bf 85}, 114032 (2012)
  [arXiv:1202.2239 [hep-ph]].


\bibitem{liangbayar}
 W.~H.~Liang, M.~Bayar and E.~Oset,
  arXiv:1610.08296 [hep-ph].


\bibitem{albertus}
  C.~Albertus, E.~Hernandez and J.~Nieves,
  Phys.\ Rev.\ D {\bf 71}, 014012 (2005)
  [nucl-th/0412006].


\bibitem{juanito}
  K.~C.~Bowler {\it et al.} [UKQCD Collaboration],
  Phys.\ Rev.\ D {\bf 57}, 6948 (1998)
  [hep-lat/9709028].

\bibitem{rocamai}
 L.~Roca, M.~Mai, E.~Oset and Ulf-G. Mei{\ss}ner,
  Eur.\ Phys.\ J.\ C {\bf 75}, no. 5, 218 (2015)
  [arXiv:1503.02936 [hep-ph]].

\bibitem{lhcbexp}
 R.~Aaij {\it et al.} [LHCb Collaboration],
  Phys.\ Rev.\ Lett.\  {\bf 115}, 072001 (2015)
  [arXiv:1507.03414 [hep-ex]].

\bibitem{sekihara}
 F.~S.~Navarra, M.~Nielsen, E.~Oset and T.~Sekihara,
  Phys.\ Rev.\ D {\bf 92}, no. 1, 014031 (2015)
  [arXiv:1501.03422 [hep-ph]].

\bibitem{sekihara2}
 T.~Sekihara and E.~Oset,
  Phys.\ Rev.\ D {\bf 92}, no. 5, 054038 (2015)
  [arXiv:1507.02026 [hep-ph]].

\bibitem{pdg}
C. Patrignani et al. (Particle Data Group), Chin.  Phys.  C {\bf 40}, 100001 (2016).


\end{thebibliography}

\end{document}